\documentclass[sigconf, anonymous=false,authorversion=true, nonacm=true]{acmart}

\setcopyright{acmlicensed}
\copyrightyear{2020}
\acmYear{2020}

\acmConference[SIGIR '20]{The 43rd International ACM SIGIR Conference on Research \& Development in Information Retrieval}{July 25--30, 2020}{Xi'an, China}
\acmBooktitle{SIGIR '20: The 43rd International ACM SIGIR Conference on Research \& Development in Information Retrieval, July 25--30, 2020, Xi'an, China}
\acmPrice{15.00}

\usepackage{xspace}
\usepackage{subcaption}
\usepackage{caption}
\usepackage{graphicx}
\usepackage{multirow}
\usepackage[show]{chato-notes}

\begin{document}

\title{Query-level Early Exit for Additive Learning-to-Rank Ensembles}

\author{Claudio Lucchese}
\affiliation{%
  \institution{Ca' Foscari University of Venice, Italy}
}
\email{claudio.lucchese@unive.it}

\author{Franco Maria Nardini}
\affiliation{%
  \institution{ISTI-CNR, Pisa, Italy}
}
\email{francomaria.nardini@isti.cnr.it}

\author{Salvatore Orlando}
\affiliation{%
  \institution{Ca' Foscari University of Venice, Italy}
}
\email{orlando@unive.it}

\author{Raffaele Perego}
\affiliation{%
  \institution{ISTI-CNR, Pisa, Italy}
}
\email{raffaele.perego@isti.cnr.it}

\author{Salvatore Trani}
\affiliation{%
  \institution{ISTI-CNR, Pisa, Italy}
}
\email{salvatore.trani@isti.cnr.it}

\renewcommand{\shortauthors}{Lucchese et al.}

\newcommand{\ltr}{\texttt{LtR}\xspace}
\newcommand{\dataset}[1]{{\sf #1}\xspace}
\newcommand{\sourcecodelocation}{\url{http://quickrank.isti.cnr.it}}

\newcommand{\algoname}[1]{{\sc #1}\xspace}
\newcommand{\istella}{\algoname{Istella}}
\newcommand{\istelladata}[1]{\dataset{{\algoname{Istella-X}}$^{\sf #1}$}}

\newcommand{\mart}{\algoname{MART}}
\newcommand{\lmart}{\algoname{$\lambda$-Mart}}

\newcommand{\etal}{\emph{et al.} }
\newcommand{\ie}{i.e., }
\newcommand{\eg}{e.g., }
\newcommand{\ndcg}{NDCG}

\begin{abstract}
Search engine ranking pipelines are commonly based on large ensembles of machine-learned decision trees. The tight constraints on query response time recently motivated researchers to investigate algorithms to make faster the traversal of the additive ensemble or to early terminate the evaluation of documents that are unlikely to be ranked among the top-$k$.
In this paper, we investigate the novel problem of \textit{query-level early exiting}, aimed at deciding the profitability of early stopping the traversal of the ranking ensemble for all the candidate documents to be scored for a query, by simply returning a ranking based on the additive scores computed by a limited portion of the ensemble. Besides the obvious advantage on query latency and throughput, we address the possible positive impact of query-level early exiting on ranking effectiveness. To this end, we study the actual contribution of incremental portions of the tree ensemble to the ranking of the top-$k$ documents scored for a given query. Our main finding is that queries exhibit different behaviors as scores are accumulated during the traversal of the ensemble and that query-level early stopping can remarkably improve ranking quality. 
We present a reproducible and comprehensive experimental evaluation, conducted on two public datasets, showing that query-level early exiting achieves an overall gain of up to 7.5\% in terms of NDCG@10 with a speedup of the scoring process of up to $2.2$x.
\end{abstract}


\maketitle

\section{Introduction}
\label{sec:intro}
Ensembles of regression trees, generated by algorithms such as LambdaMART \cite{lambdamart,burges2010ranknet} and Gradient Boosted Regression Trees (GBRT) \cite{Friedman00greedyfunction}, are  widely considered the state-of-the-art learning-to-rank solutions, currently employed in modern Web search engines to effectively re-rank the set of documents selected for each user query, and to finally return the most relevant ones among billions of indexed documents. However, while Web search is a scenario characterized by tight query processing latency constraints,  these complex machine-learned ranking models are computationally demanding. 
learning-to-rank solutions, and as a result, 
These two contrasting aspects lead to the investigation of techniques focused on speeding-up the exploitation of tree ensembles without hindering their effectiveness. 
Some techniques have been proposed aimed to improve performance through novel tree traversal \cite{rapidscorer18,Dato2016,quickscorer15}, ensemble pruning \cite{lucchese2017x,xcleaver,Lucchese2016}, and budget-aware learning algorithms \cite{Wang2010,asadi2013training}. Other recent studies focused on balancing feature costs (efficiency) and system effectiveness across multiple re-ranking stages using cascading loss functions \cite{RoiSIGIR17}. Another approach to improve efficiency in multi-stage retrieval regards the minimization of the number of documents that must be re-ranked for each query \cite{relyahoo,Capannini20161161}. 

In this paper, we investigate the novel problem of \emph{query-level early exiting}, by studying the profitability of early terminating the traversal of the ranking ensemble for all the candidate documents retrieved for a query. For the early-terminated queries we simply return a ranked list based on the partial scores computed until the exit trees. Cambazoglu \emph{et al.}~\cite{EarlyBarla10} already introduced early exit strategies for additive ensembles of regression trees, but their technique selectively stops the ensemble traversal at document-level, rather than at query-level.
Their document-level early exit strategies can speedup the score computation with almost no loss in effectiveness. Our approach substantially differs from  this proposal, because our early exit decision does not apply to a single document, but to the whole set of candidate documents retrieved for a given query. 
We investigate how different portions of the tree ensemble contribute to the ranking of the top-$k$ documents and we experimentally show that our approach has: \emph{i)} an obvious advantage on query latency and throughput, due to the time saved by avoiding the traversal of large portions of the ensemble, and \emph{ii)} a positive impact  on ranking effectiveness, due to the early termination of queries on which the whole ensemble performs worse than a portion of it.
\begin{figure}[b!]
    \centering
    \vspace{-4mm}
    \includegraphics[width=\linewidth]{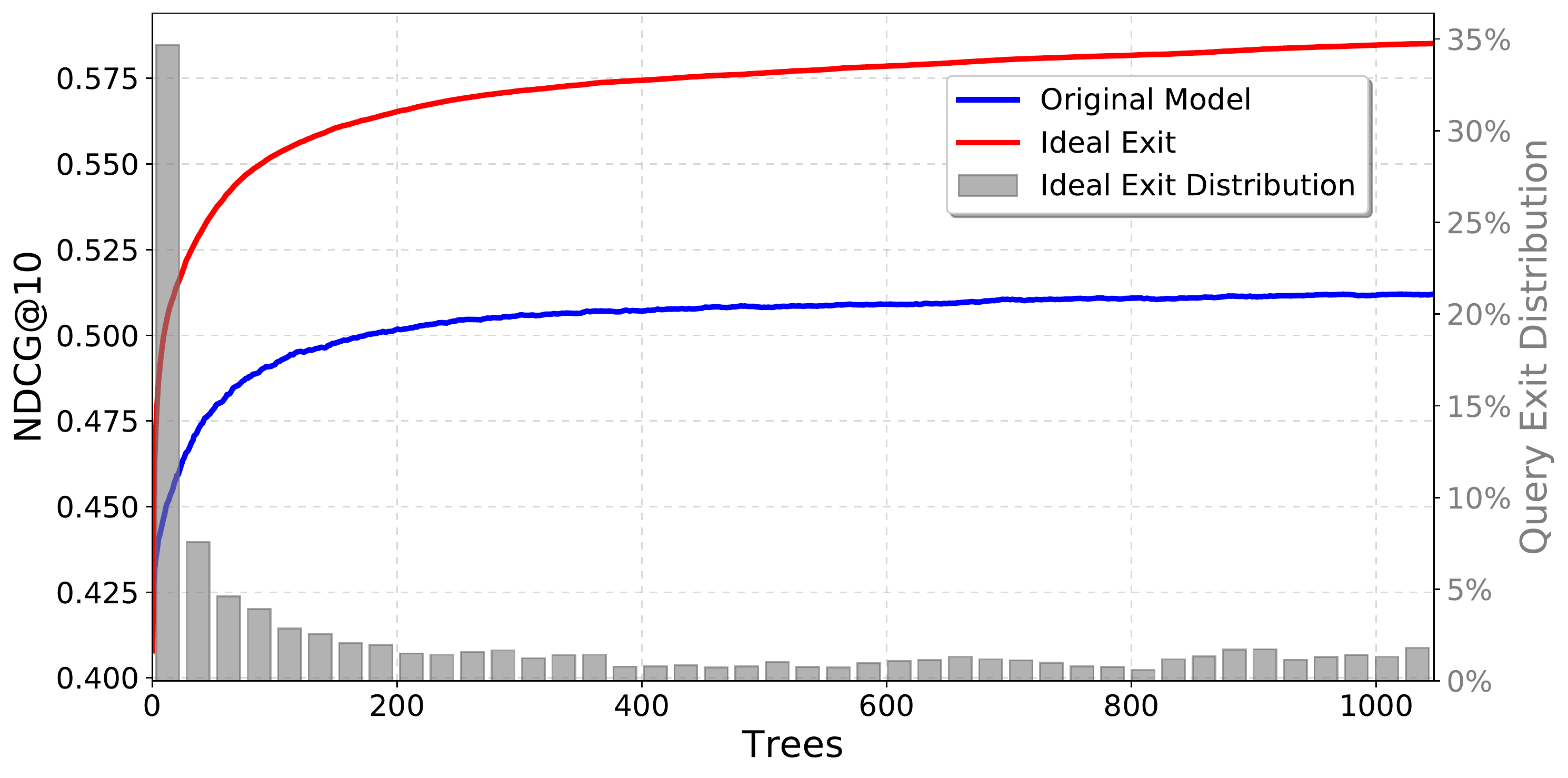}
    \vspace{-7mm}
    \caption{Ideal early exit on MSLR-WEB30K.}
    \label{fig:optimal_exit}
\end{figure}

\begin{figure*}[t!]
        \begin{subfigure}[b]{0.33\textwidth}
                \centering
                \includegraphics[width=.95\linewidth]{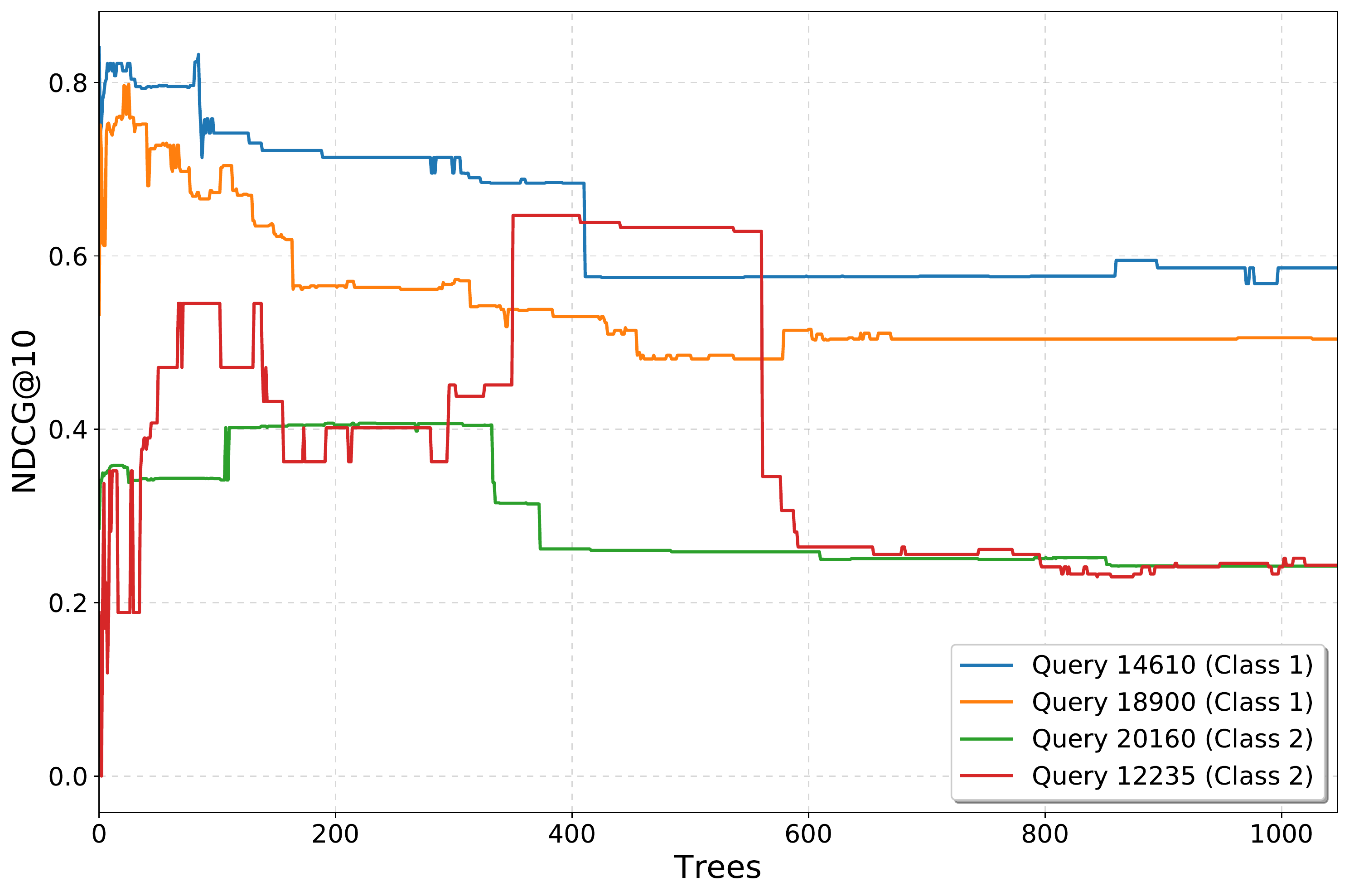}
                \vspace{-2mm}
                \caption{Worsening Queries}
                \label{fig:query-classes-12}
        \end{subfigure}%
        \begin{subfigure}[b]{0.33\textwidth}
                \centering
                \includegraphics[width=.95\linewidth]{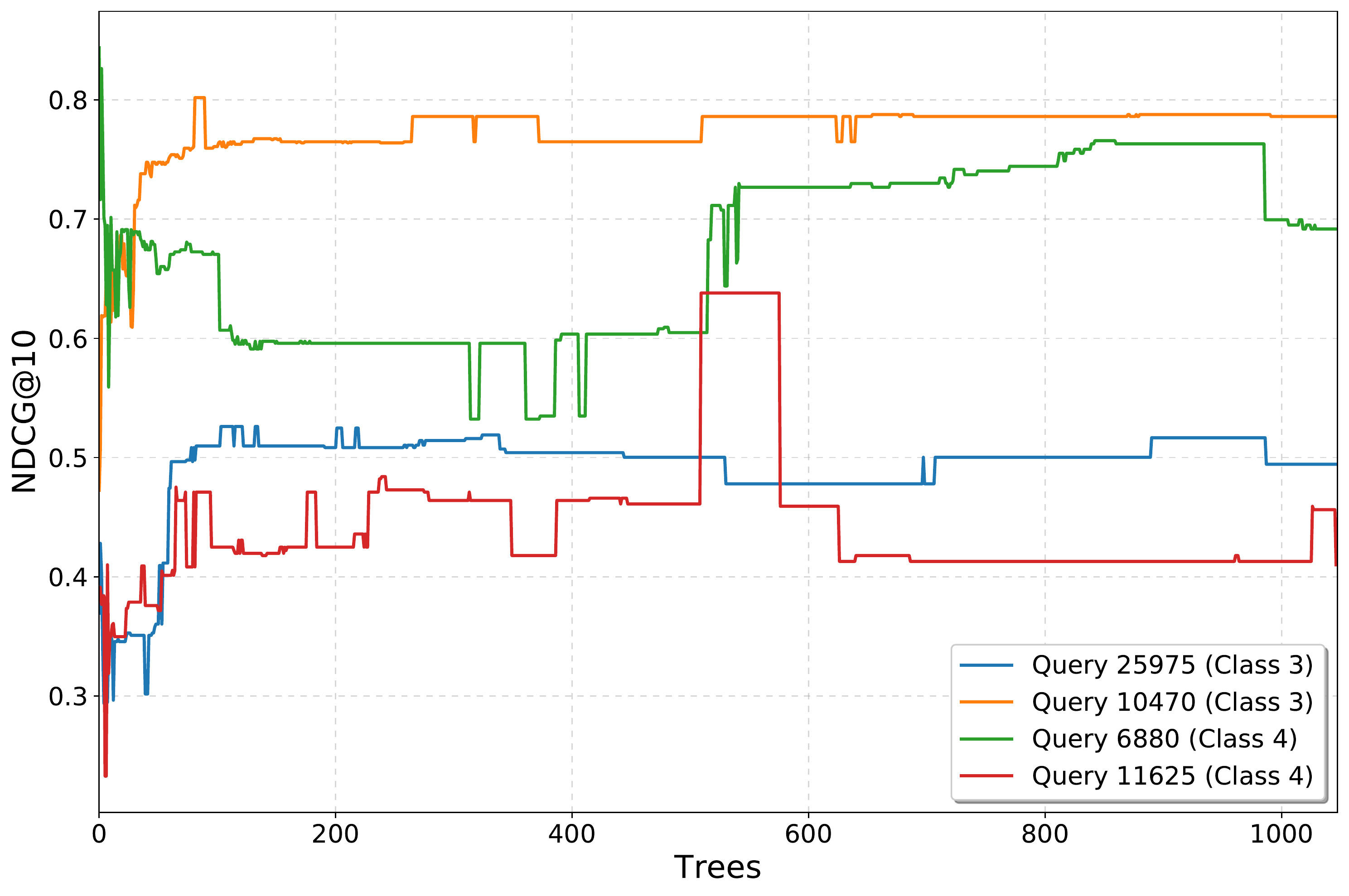}
                \vspace{-2mm}
                \caption{Flat Queries}
                \label{fig:query-classes-34}
        \end{subfigure}%
        \begin{subfigure}[b]{0.33\textwidth}
                \centering
                \includegraphics[width=.95\linewidth]{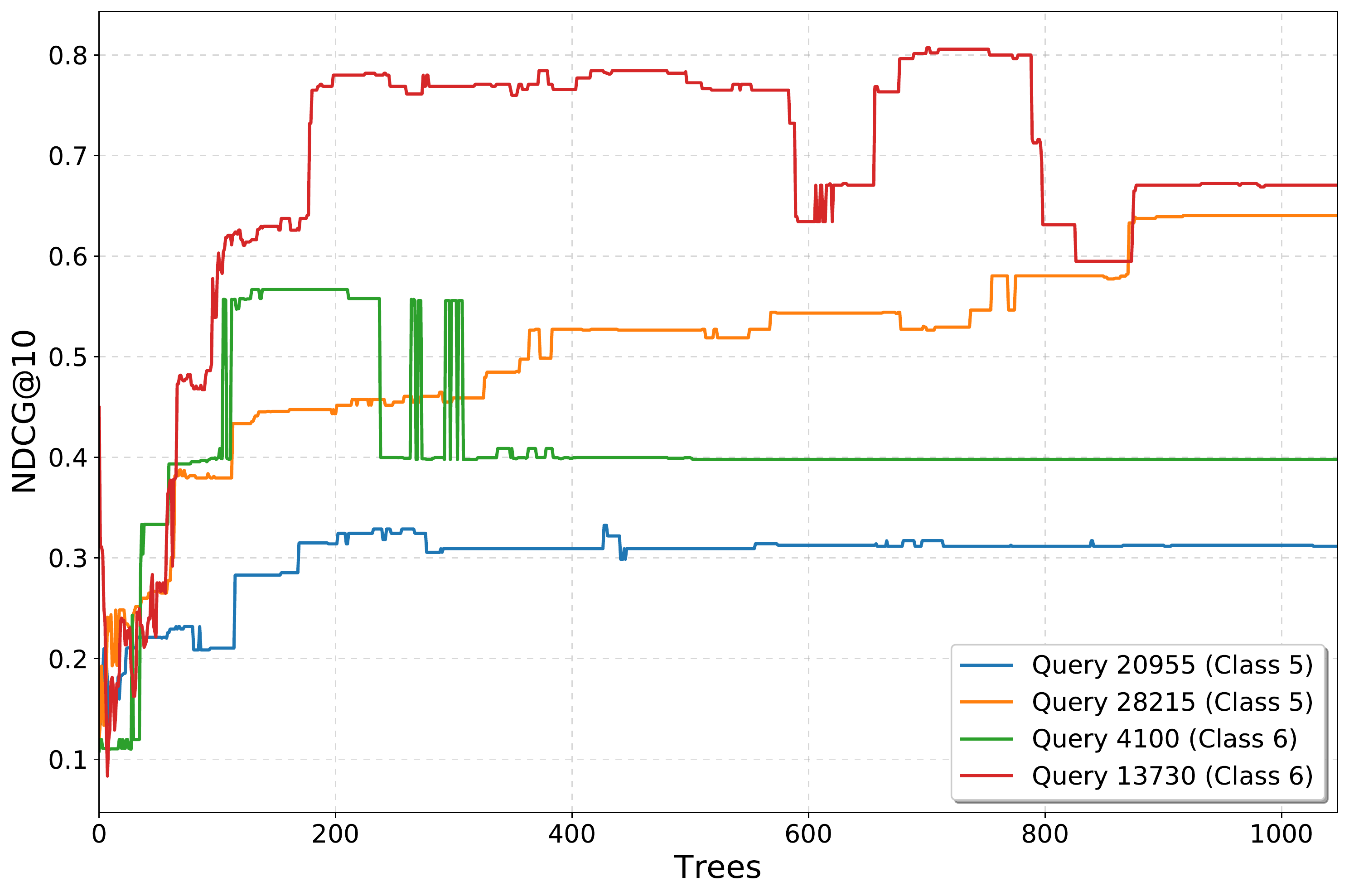}
                \vspace{-2mm}
                \caption{Improving Queries}
                \label{fig:query-classes-56}
        \end{subfigure}%
        \vspace{-2mm}
        \caption{NDCG@10 of different queries by varying the number of trees of the forest on the MSLR-WEB30K dataset, test split.}
        \label{fig:query-classes}
\end{figure*}

To concretely quantify the potential impact of query-level early exiting on the ranking effectiveness, we assume to have an ``oracle'' that, given a ranking ensemble and a set of candidate documents retrieved for a query, is able to predict the ``best''  exit point in the ensemble for the query, i.e., the portion of the whole ensemble providing the best value for the quality metrics of interest.
Figure \ref{fig:optimal_exit} plots the ranking performance measured in terms of average NDCG@10 for an ensemble of $1,047$ trees, where either an ideal query-based early exiting strategy using the oracle is employed (red line), or all queries traverse the whole ensemble (blue line).
The ensemble is a LambdaMART model learned on the MSLR-WEB30K\footnote{\url{https://www.microsoft.com/en-us/research/project/mslr/}} dataset (fold 1) using LightGBM\footnote{\url{https://github.com/Microsoft/LightGBM}} as a training algorithm. The plots refer to the performance obtained by scoring the associated test split. 
Of course, this perfect oracle is not available at query time and the top red line shown in the plot is the upper-bound of any query-level early exiting strategy. However, from the figure we can see that the effectiveness gain achievable with such a technique is remarkable, i.e., more than 7 points of NDCG@10 (+14\%). Figure \ref{fig:optimal_exit} also reports 
the distribution of the queries stopping at each tree in the ensemble (see the right Y-axis), as determined by our ideal early exiting strategy. We can observe that this distribution is very skewed, i.e., that most ideal exit points are concentrated at the beginning of the ensemble, and a relatively small portion of the queries need to complete the ensemble traversal to maximize NDCG@10. This may have a potential large impact on query latency and throughput, since many queries can be early terminated soon, by saving the time required to traverse the largest part of the tree ensemble.

\section{Query-level early exit}
In this section we study how the NDCG@10 changes as the scores of documents are incrementally accumulated through the different portions of the ensemble. Figure \ref{fig:query-classes} reports the result of this exploratory data analysis, still conducted on the test set of the MSLR-WEB30K dataset (fold 1). 
Each plot of the Figure shows the NDCG@10 (Y-axis) of four example queries as a function of the number of trees of the ensemble traversed (X-axis). We are able to identify six main classes of queries, that we represent schematically with the behavior of the example queries. %
Queries in Classes 1 and 2 belong to the category of the  \textit{worsening queries}, whose NDCG@10 tends to decrease as we approach the end of the ensemble (see \ref{fig:query-classes}.a). Queries of Class 1 exhibits a decreasing NDCG@10 when we increase the portion of the ensemble traversed by its documents (e.g., query 14610 loses more than 20 points). Queries of Class 2 are instead characterized by an NDCG@10 that increases until a maximum, and then starts decreasing falling below the starting point (e.g., query 12235).
Classes 3 and 4 are grouped together in the category of \textit{flat queries} in the plots of Figure \ref{fig:query-classes}.b. Class 3 includes all the queries with no significant changes in the NDCG@10 through the whole forest, e.g., query 25975. Similarly, Class 4 groups all the queries with no significant change in  NDCG@10, except of some local variations of the metric in some region of the ensemble. Finally, Classes 5 and 6 belong to the category of \textit{improving queries}, represented in Figures \ref{fig:query-classes}.c. In particular, queries in Class 5 shows a increasing NDCG@10 when the scoring proceed through the forest (e.g., query 20955), while queries in Class 6 show a similar behavior reaching a maximum NDCG@10 value and then from a specific point in the ensemble on start to decrease (e.g., query 4100). These queries differ from Class 2 queries in the positive delta in terms of effectiveness showed at the end of the forest compared to the first part.
The analysis reported in Figure \ref{fig:query-classes} show that there is a significant margin of improvement in terms of NDCG@10 if we are able to effectively classify queries belonging to the different classes and to early terminate the queries in the classes 1, 2, 4, and 6 before their performance starts to decrease. 
Besides the obvious improvement in query throughput this performance improvement motivates our investigation of a query-level early exiting strategy.

\subsection{Impact of Query-level Early Exit}
We now deepen the analysis of query early exit, by introducing the concept of ``sentinel'', i.e., a specific tree of the ensemble where the scores of the documents belonging to a query are accumulated and a decision about early stopping query evaluation is taken. 
Although we still use an oracle for deciding about  query early, reducing the number of sentinel points where the decisions are taken, makes our technique more practicable in an real ranking system.
In Section~\ref{sec:development} we discuss in more detail this point, and observe that tree ensembles are usually processed in blocks~\cite{block-scorer,IEEETPDS}, and block boundaries are the best candidate points where sentinels can be placed. 
We choose the best assignment of a given number of sentinels with an exhaustive search algorithm testing all the combinations of positions the sentinels can take and measuring  on the validation set the resulting  impact in terms of  NDCG@10. 
Finally, the positions of the sentinels maximizing  the average NDCG@10 are chosen. 
Obviously, by increasing the number of sentinels, we also increase the overall average NDCG@10. To this regards, Figure~\ref{fig:optimal_exit} is a special case computed using a distinct sentinel for each tree of the ensemble. However, in the following we limit our study to the case of only \textit{two sentinels}, placed at discrete points of the ensemble, indeed at discrete positions multiple of 25 trees.  We also study the special case of a further sentinel placed after the first tree, aimed to capture the spike of the early exit query distribution of Figure~\ref{fig:optimal_exit}, occurring at the very beginning of the ensemble.

\begin{figure}[t]
    \centering
    \includegraphics[width=0.9\linewidth]{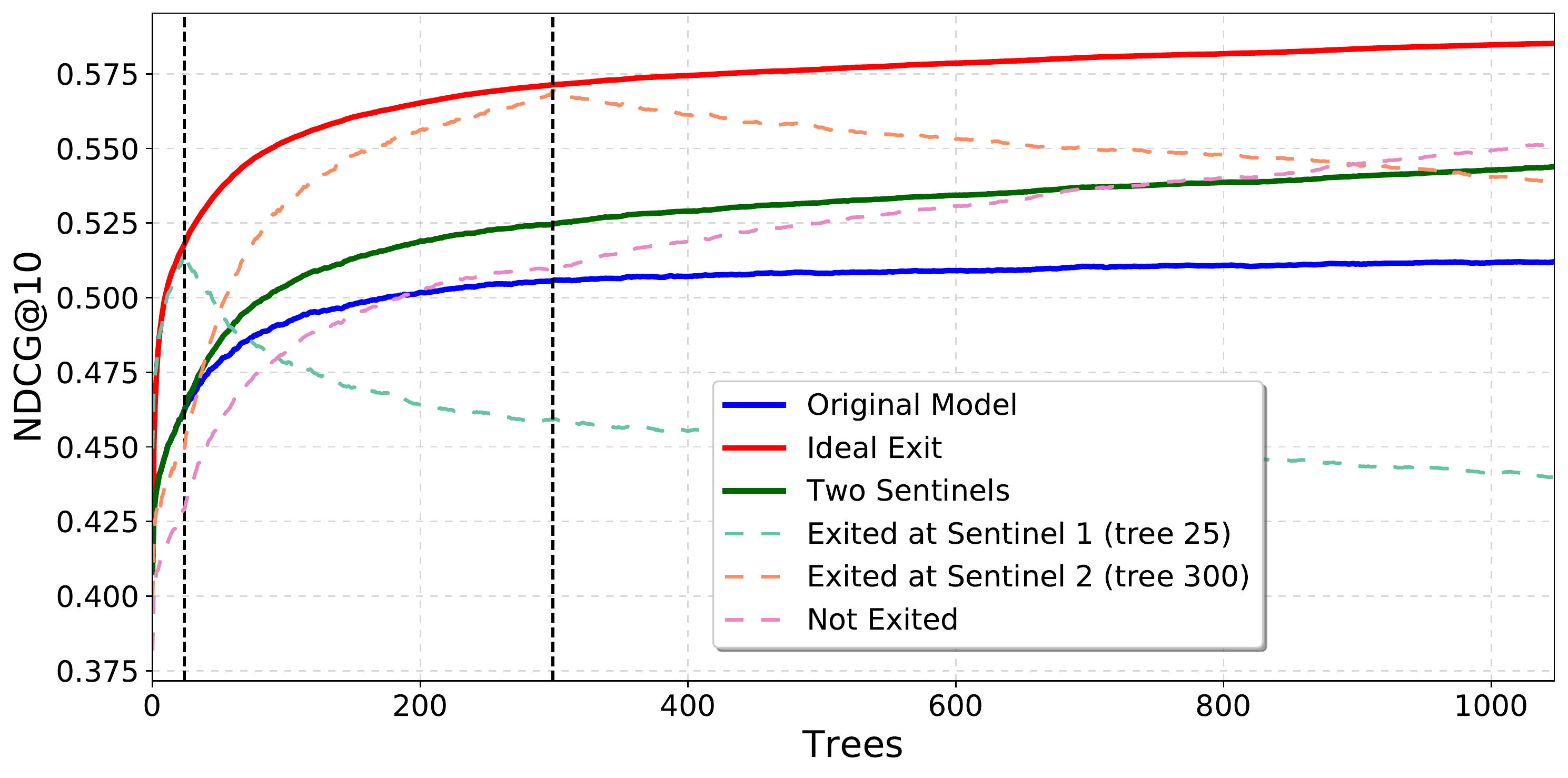}
    \vspace{-4mm}
    \caption{Query-level early exit with two sentinels on MSLR-WEB30K.\label{foptimally_placed}}
    \vspace{-5mm}
\end{figure}

\smallskip
\noindent \textbf{Analysis on MSLR-WEB30K}.
We analyze the placement of only two sentinels when scoring the test set of MSLR-WEB30K (fold 1) with the same LambdaMART model of Figure \ref{fig:optimal_exit}. The best placements of the sentinels are the $25^{th}$ and the $300^{th}$ trees. Figure \ref{foptimally_placed} reports an analysis of the performance observed, where the sentinels are shown as vertical dashed black lines. 
We report the average NDCG@10 (Y-axis) of the original and the ideal models as blue and red solid lines, respectively. We also report the average NDCG@10 when employing the two-sentinels technique as a green solid line. The contribution of the two sentinels is relevant to the average NDCG@10, because the green line is significantly above the blue one (original model) after the first sentinel and, even more, after the second sentinel. This is due to a significant gain of the queries early stopped by the two sentinels. 
The effect of the two sentinels is thus apparent in terms of final effectiveness with an increment of the average NDCG@10 of more than 3 points. 
Figure \ref{foptimally_placed} also reports the average NDCG@10 of the queries belonging to the three groups of queries, i.e., queries early stopped by the first and the second sentinel, and the remaining queries that traverse the full ensemble. Indeed, early-stopped queries shows the maximum average NDCG@10 in correspondence of the sentinel selected. More interestingly, the average performance of these queries would drop significantly in case they were not stopped. As an example, the queries early stopped by the first sentinel drop by more than 7 points in terms of average NDCG@10, while the drop is limited to only 3 points for queries early stopped by the second sentinel. On the other hand, queries that continue to be processes are characterized by a monotonically increasing average NDCG@10. Another interesting experimental evidence that is worth highlighting regards the behaviour of the average NDCG@10 of the ideal model (red line in Figure \ref{foptimally_placed}) at the very early stages. The red line, in fact, is almost vertical in the first trees of the ensemble. This means that the important fraction of the queries that is stopped very early contributed significantly to the average NDCG@10. Figure \ref{foptimally_placed} thus confirms that the introduction of query-based early exit with only two sentinels is already beneficial for the overall ranking quality of the system (green solid line), although worse than the ideal case with sentinels at each tree (red solid line).

\begin{table}[htb]
    \def\arraystretch{1.1}
    \setlength{\tabcolsep}{1.3mm}
    \centering
    \caption{Two sentinels on  MSLR-WEB30K.\label{tab:perf_msn}}
    \vspace{-3mm}
    \begin{tabular}{l|c|c|l|l}
    \multirow{2}{*}{\# sentinel} & \multirow{2}{*}{\# queries} & \multicolumn{2}{c|}{NDCG@10} & \multirow{2}{*}{speedup} \\
    \cline{3-4} & & L & \multicolumn{1}{c|}{sentinel} & \\
    \toprule
    1 @ tree=25 & $2,024_{(33\%)}$ & $0.4399$ & $0.5161_{(+17.3\%)}$ & $41.9$x \\
    2 @ tree=300 & $1,339_{(22\%)}$ & $0.5391$ & $0.5694_{(+5.6\%)}$ & $3.5$x \\
    L @ tree=1,047 & $2,754_{(45\%)}$ & $0.5518$ & $0.5518_{(+0\%)}$ & $1$x \\
    \midrule
    Overall & $6,117_{(100\%)}$ & $0.5120$ & $0.5439_{(+6.2\%)}$ & $1.9$x \\
    \bottomrule
    \end{tabular}
    \vspace{-2mm}
\end{table}

Table \ref{tab:perf_msn} reports the detailed figures, i.e., the number of queries, the average NDCG@10 and the speedup achieved by the two sentinels introduced in the model. The first two rows of the table report the contribution of each sentinel, while the third row reports the performance achieved by the fraction of queries not early stopped. We also summarize the overall performance on the test set computed as a weighted mean of the results for the three groups of query. The first sentinel is responsible for early exiting $2,024$ queries (33\% of the test set). By doing so, these queries achieve an average NDCG@10 of $0.5161$, while their performance drops to $0.4399$ when no query-level early termination is adopted. Therefore, for this group of queries, our proposed technique improves the NDCG@10 by $17.3$\%. This performance is achieved by using only the first $25$ trees of the forest instead of $1,047$. This also leads to a significant speedup of up to $41.9$x of the scoring time, considering that this time is linearly proportional to the number of queries (associated documents) and the tree actually traversed. The second sentinel early stops $1,339$ queries (22\% of the test set), achieving an average NDCG@10 of $0.5688$. For this group of queries, the improvement with respect to the full model is $5.6$\% as the NDCG@10 achieved with no early termination is $0.5391$. Moreover, the second sentinel allows a speedup of $3.5$x. The remaining queries are not early stopped. For this group of queries no performance gain and no speedup is achieved. Overall, our new query-level early stopping allows for an improvement of $6.2$\% in terms of average NDCG@10 while achieving a speedup of $1.9$x, thus halving the average time needed by the system to score a query.

\begin{table}[b!]
    \def\arraystretch{1.1}
    \setlength{\tabcolsep}{1.3mm}
    \vspace{-2mm}
    \caption{Three sentinels on MSLR-WEB30K.\label{tab:perf_msn_first}}
    \vspace{-3mm}
    \centering
    \begin{tabular}{l|r|c|l|l}
    \multirow{2}{*}{\# sentinel} & \multirow{2}{*}{\# queries} & \multicolumn{2}{c|}{NDCG@10} & \multirow{2}{*}{speedup} \\
    \cline{3-4} & & L & \multicolumn{1}{c|}{sentinel} & \\
    \toprule
    1 @ tree=1 & $1,605_{(26\%)}$ & $0.4120$ & $0.5043_{(+22.4\%)}$ & $1,047$x \\
    2 @ tree=25 & $901_{(15\%)}$ & $0.4933$ & $0.5509_{(+11.7\%)}$ & $41.9$x \\
    3 @ tree=300 & $1,157_{(19\%)}$ & $0.5497$ & $0.5795_{(+5.4\%)}$ & $3.5$x \\
    L @ tree=1,047 & $2,454_{(40\%)}$ & $0.5664$ & $0.5664_{(+0\%)}$ & $1$x \\
    \midrule
    Overall & $6,117_{(100\%)}$ & $0.5120$ & $0.5503_{(+7.5\%)}$ & $2.2$x \\
    \bottomrule
    \end{tabular}
\end{table}

\smallskip
\noindent \textbf{Analysis on MSLR-WEB30K with three sentinels}.
In the previous analysis we point out that the ideal model (red solid line in Figure \ref{foptimally_placed}) shows a significant improvement of the average NDCG@10 very early in the ensemble. Thus, we further deepen the analysis conducted for MSLR-WEB30K by testing the performance achieved when placing a further sentinel after the first tree of the ensemble, while the others are still optimally placed at the $25^{th}$ and $300^{th}$ trees. Table \ref{tab:perf_msn_first} reports the results achieved. The first sentinel (tree no. 1) is able to early stop $1,605$ queries (26\%) with a gain of average NDCG@10 of more than 9 points (22.4\%), and a speedup of $1,047$x as only one tree of the forest is employed to score this group of queries. Indeed, the second sentinel (tree no. 25) now early stops $901$ queries (15\% of the test set) with an improvement of NDCG@10 of almost 6 points (11.7\%) and a speedup of $41.9$x. It is worth remarking that the first two sentinels placed on tree no. 1 and 25 of the ensemble allows to early stop 41\% of the queries while, in the case of two sentinels, the first one only stops 33\% of the queries. The third sentinel (tree no. 300) now achieves $5.4$\% of improvement with a speedup of $3.5$x  on $19$\% of the queries. Indeed, only the 40\% of the queries are scored using the entire ensemble. Overall, the improvement of NDCG@10 is $7.5$\% with a speedup of $2.2$x. Although this is an interesting result, the placement of the sentinel on the first tree is challenging for the practical development of an effective sentinel classifier (see Section \ref{sec:development}). 

\smallskip
\noindent \textbf{Analysis on Istella-S}.
We perform the same analysis on a second dataset, i.e., the Istella-S dataset\footnote{\url{http://quickrank.isti.cnr.it/istella-dataset/}} by employing the same methodology used for MSLR-WEB30K. For this dataset, the original LambdaMART model consists of $1,304$ trees and the assignment of the sentinels accordingly to the exhaustive search algorithm is in correspondence of the trees no. 25 and 450. The experimental results, reported in Table \ref{tab:perf_istella}, are consistent with the ones obtained on the MSLR-WEB30K dataset. Indeed, in this case the gain in terms of average NDCG@10 achieved by the two sentinels ranges from $4.1$\% to $9.0$\% with a speedup ranging from $2.9$x to $52.2$x. Overall, the two sentinels achieves a gain of $3.4$\% with a speedup of $1.8$x. 

\begin{table}[t!]
    \def\arraystretch{1.1}
    \setlength{\tabcolsep}{1.3mm}
    \centering
    \caption{Two sentinels on  Istella.\label{tab:perf_istella}}
    \vspace{-3mm}
    \begin{tabular}{l|c|c|l|l}
    \multirow{2}{*}{\# sentinel} & \multirow{2}{*}{\# queries} & \multicolumn{2}{c|}{NDCG@10} & \multirow{2}{*}{speedup} \\
    \cline{3-4} & & L & \multicolumn{1}{c|}{sentinel} & \\
    \toprule
    1 @ tree=25 & $1,928_{(30\%)}$ & $0.7250$ & $0.7904_{(+9.0\%)}$ & $52.2$x \\
    2 @ tree=450 & $1,447_{(22\%)}$ & $0.7654$ & $0.7968_{(+4.1\%)}$ & $2.9$x \\
    L @ tree=1,304 & $3,153_{(48\%)}$ & $0.8052$ & $0.8052_{(+0\%)}$ & $1$x \\
    \midrule
    Overall & $6,528_{(100\%)}$ & $0.7727$ & $0.7990_{(+3.4\%)}$ & $1.8$x \\
    \bottomrule
    \end{tabular}
    \vspace{-3mm}
\end{table}

\section{Towards early exit classifiers}
\label{sec:development}
Our study shows the potential of a query-level early exit strategy. To be employed in real-world query processors, this technique requires the deployment of binary classifiers, one for each sentinel in the ensemble aiming at identifying queries that can be safely early stopped.
To this end, we note that the integration of classifiers in the scoring pipeline is compatible with the data model of modern scoring algorithms \cite{block-scorer,IEEETPDS}, that work by partitioning the ensemble in blocks to reduce the memory footprint of the data structures and to better exploit the memory hierarchies.
In this settings, each block of trees is used to score the set of candidate documents retrieved for a given query. The final score of each document is the sum of the scores produced by the different blocks. Block boundaries are thus the candidate points where sentinels enabling query-level early exit can be placed. 
The most important point to address in the development of this kind of classifier regards the features to use to effectively identify the queries to early stop while scoring them. Indeed, this point poses interesting challenges for two main reasons: i) as the classifier works online with the scoring of the query, the computation/extraction of the features should be fast to fully exploit the speedup achieved by our technique, ii) the features should be powerful enough to avoid type I (false positive) and II (false negative) misclassification errors, with a priority on the former as  wrongly early stopped queries might result in poor ranking quality. An important contribution in that direction could be provided by listwise features, e.g., aggregations of the top-$k$ document scores, and their trends on several consecutive trees that can be used as features to model the decision of early stopping the query. The unavailability of this kind of features for the first tree forces the usage of single document scores coupled with statistical signals to develop an effective sentinel classifier.

\section{Conclusions and Future Work}
\label{sec:conclusions}
We investigated the novel problem of \textit{query-level early exit}, aimed at improving query processing throughput and the ranking quality 
by early stopping the traversal of a ranking ensemble for all the documents to be scored for a query. Interestingly, we experimentally assessed a large possible impact on ranking effectiveness, due to many queries answered more effectively by a fraction of the ranking ensemble rather than the whole model.  In case of two sentinels, where an oracle decides about early-exiting a query, 
we measured an ideal  gain  of  $7.5$\% points in  NDCG@10, and an overall speedup of  $2.2$x for the scoring process. Future research stimulated by our analysis will assess if such large opportunities for improvement can be actually exploited by suitable classifiers placed in correspondence of specific sentinels.

\section{Acknowledgements}
Work partially supported by the BIGDATAGRAPES project funded by the EU Horizon 2020 research and innovation programme under grant agreement No. 780751, and by the OK-INSAID project funded by the Italian Ministry of Education and Research (MIUR) under grant agreement No. ARS01\_00917.

\bibliographystyle{ACM-Reference-Format}



\end{document}